# A generalized neutral theory explains static and dynamic properties of biotic communities

Michael Kalyuzhny[1]*, Ronen Kadmon[1] & Nadav M. Shnerb[2]

[1]Department of Evolution, Ecology and Behavior, Institute of Life Sciences, Hebrew University of Jerusalem, Givat-Ram, Jerusalem 91904, Israel.

[2]Department of Physics, Bar-Ilan University, Ramat Gan 52900, Israel.

*Correspondence to: michael.kalyuzhny@mail.huji.ac.il.

**Abstract**

**Understanding the forces shaping ecological communities is crucially important to basic science and conservation. In recent years, considerable progress was made in explaining communities using simple and general models, with neutral theory as a prominent example. However, while successful in explaining static patterns such as species abundance distributions, the neutral theory was criticized for making unrealistic predictions of fundamental dynamic patterns. Here we incorporate environmental stochasticity into the neutral framework, and show that the resulting generalized neutral theory is capable of predicting realistic patterns of both population and community dynamics. Applying the theory to real data (the tropical forest of Barro Colorado Island), we find that it better fits the observed distribution of short-term fluctuations, the temporal scaling of such fluctuations, and the decay of compositional similarity with time, than the original theory, while retaining its power to explain static patterns of species abundance. Importantly, although the proposed theory is neutral (all species are functionally equivalent) and stochastic, it is a niche-based theory in the sense that species differ in their demographic**

**responses to environmental variation. Our results show that this integration of niche forces and stochasticity within a minimalistic neutral framework is highly successful in explaining fundamental static and dynamic characteristics of ecological communities.**

Understanding the dynamic processes that govern biotic communities is among the greatest challenges in ecology. Traditionally, most ecological thinking was devoted to the identification of deterministic mechanisms such as functional trade-offs and species-specific interactions that stabilize communities and generate equilibrial species composition [*1*]. This approach requires the estimation of a large number of parameters [*2*], a formidable task given the complexity of ecological interactions.

A different approach, pioneered by MacArthur and Wilson's theory of island biogeography, adopted a minimalistic description of ecological systems, trying to capture essential features of communities by models where all species are functionally equivalent ('neutral'). This approach neglects the details of interspecies interactions, and is therefore much less demanding in terms of the number of parameters required to describe community dynamics. As opposed to species-specific interactions, the neutral approach emphasizes the role of stochastic processes as the only pattern generating mechanisms.

The neutral approach has gained particular interest since the presentation of the Unified Neutral Theory of Biodiversity (hence UNTB) by Hubbell [*3*]. The main success of UNTB is its ability to account for empirically observed species abundance distributions [*3-5*] (SAD) with a simple model in which all species are demographically equivalent and their dynamics are governed solely by a stochastic birth-death-migration process (demographic stochasticity). Following the empirical success and parsimony of UNTB, theories are currently being developed to integrate this purely stochastic framework with deterministic forces [*6, 7*].

While UNTB is successful at explaining a number of static biodiversity patterns, there is growing evidence that it fails to account for the long- (evolutionary) and short- (ecological) term dynamics of populations and communities. At the evolutionary time scale, extinction times of common species appear to be unrealistically long (up to billions of years), while those of rare species appear to be too short [*8, 9*]. At the ecological time scale, empirically observed fluctuations in abundance are usually too large to be explained by UNTB [*10-13*], and they tend to be directional, i.e., have some degree of temporal correlation [*14*], which contrasts the prediction of UNTB. Moreover, UNTB cannot account for the scaling of fluctuations variance with population size: while UNTB predicts a linear scaling, empirical analyses show a prevalence of super-linear dependence [*10, 11, 15*].

Here we explore the possibility that these limitations may be resolved by incorporating fluctuations in species fitness (environmental stochasticity [*16*]) into the neutral framework. As in UNTB, our 'neutral niche theory' can be formulated at local or regional scales. At the local scale, we consider a community of $J$ individuals belonging to S species, each having fitness $f_i$ and abundance $N_i$ ($i=1,…,S$), that receives immigrants from a regional species pool ('metacommunity' in Hubbell's terminology) with a log-series distribution of species abundance. In an elementary time step one individual, chosen at random from the community, dies, and replaced by an immigrant from the regional pool with probability $m$ or an offspring of a local individual with probability 1-$m$. In the latter case, the probability that this offspring belongs to a species $k$ is:

$$P(k) = \frac{N_k f_k}{\sum_{i=1}^{S} N_i f_i} \quad (1)$$

P($k$) incorporates both the abundance and the fitness of all species; if all species have the same fitness, the dynamics is reduced to that of the UNTB.

Environmental stochasticity is introduced through species-level fluctuations in fitness. For any species, every $\tau$ time steps the fitness is redrawn (independently of the other species) from a lognormal distribution with mean 1 and variance $A$, causing periods of population decline and increase, but leaving species neutral in the long run. Thus, $\tau$ is a measure of the temporal scale of autocorrelation in environmental conditions. Recent analyses indicate that such a model of temporal fluctuations is a reasonable assumption [*2*]. The dynamics of a regional community with $J_m$ individuals is modeled using a similar approach by incorporating point speciation at rate $\mu$ instead of immigration.

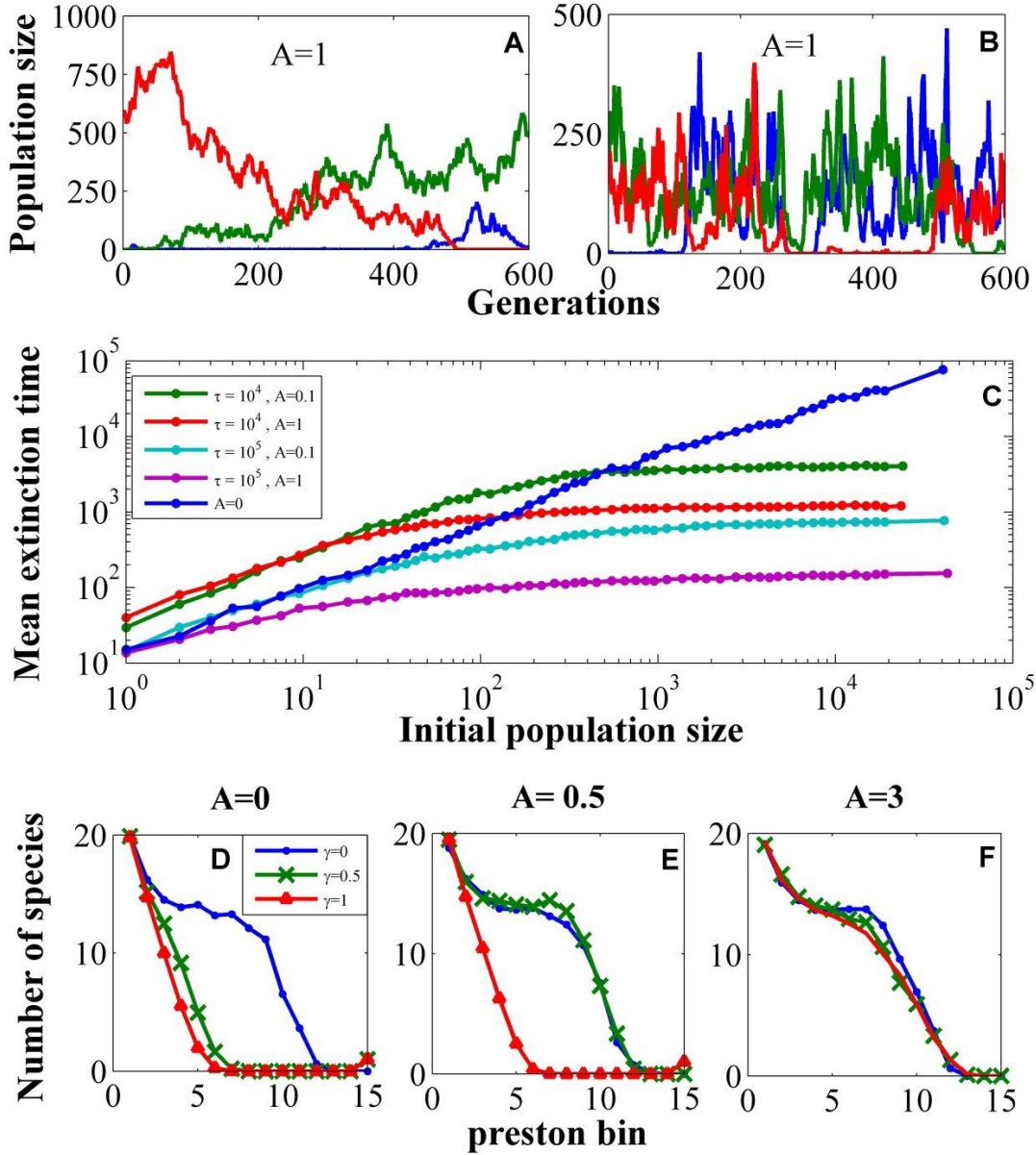

*Figure 1 **Analysis of the neutral niche model (NNTB) demonstrating the effect of environmental stochasticity on coexistence, extinction time and robustness to fitness differences.** Note that predictions of NNTB with A=0 (absence of environmental stochasticity) are analytically equivalent to UNTB. **A** and **B** compare the population dynamics of three species in the absence (**A**) and presence (**B**) of environmental stochasticity. **C** shows the dependence of mean time to extinction on initial population size at the absence of environmental stochasticity (A=0, blue curve), and under different regimes of environmental stochasticity as defined by the scale of temporal autocorrelation in environmental conditions (τ) and the variance of fitness*

*fluctuations (A). **D-F** show typical SADs at the absence (A=0) and presence (A=0.5 and A=3) of environmental stochasticity for three levels of competitive advantage (γ) of a focal species on all other species. (γ=0 implies all species are competitively equivalent). Note that in **D** both curves with γ>0 have a single species in the 15$^{th}$ bin. (see supporting materials 1 for more details).*

Importantly, this modeling approach is still neutral in the sense that all species are equivalent in the long-term (have the same time-averaged fitness)[*17*], but at the same time it can be viewed as a niche theory, since species differ from each other in their demographic responses to environmental variation. We therefore term it a Neutral Niche Theory of Biodiversity (NNTB).

Our theoretical analysis shows that NNTB overcomes the time-to-extinction problem of UNTB and its unrealistic demand for perfect neutrality (Fig. 1). One noticed result is that introducing environmental stochasticity stabilizes the dynamics of the community (Fig. 1 A, B). We attribute this result to a 'storage effect' [*18, 19*]: small populations suffer a small decrease in abundance during unfavorable conditions but can increase rapidly under favorable conditions, while large populations are subject to opposite forces. This well-known [*20*] stabilization mechanism has a profound effect on the mean time to extinction (fig. 1C): while under UNTB (A=0) the time to extinction in the regional community grows approximately linearly with initial population size [*8*], the storage effect causes a clear saturation in extinction times by generating density dependent attraction of all species to size *J/S* (fig. 1B). This saturation reflects the fact that under NNTB, populations fluctuate (widely!) around the equilibrium, with their initial size playing a minor role compared to the characteristic time of reaching extinction from equilibrium [*21*]. Moreover, the attraction to equilibrium induced by the storage effect may considerably enhance the longevity of small populations (Fig. 1C, see also Supplementary Figure 1, supporting materials 2 and Rosindell et. al. [*9*] for another resolution of this problem). Thus, incorporating environmental stochasticity in neutral theory results in more realistic predictions of extinction times for both common and rare species.

Another important consequence of the storage effect is a reduction in the sensitivity of the predicted SAD to differences in mean fitness. Previous studies have noted that under UNTB, even small differences in fitness lead to a collapse of the community into dominance by a single species [*22, 23*]. In Figure 1D-F we evaluate the sensitivity of NNTB to fitness differences by analyzing cases where a single species has different levels of competitive advantage $\gamma$ (expressed in terms of mean fitness) over all other species. As expected, under pure demographic stochasticity (A=0, Figure 1D), any competitive advantage leads to a collapse of the SAD into a scenario where the superior species dominates the system with a relatively small number of rare species. Environmental stochasticity (A>0, Figure 1E-F) generates storage effect that buffers this advantage, allowing coexistence of many species, and the stronger the stochasticity, the higher the competitive advantage that can be buffered.

We conclude that adding environmental stochasticity to the neutral framework may help to resolve two limitations of the theory – biased predictions of times to extinction [*8*] and extreme sensitivity to small differences in fitness [*22, 23*]. Moreover, while UNTB predicts only one type of SAD in the regional community (log-series), our extended model can account for all commonly observed SADs [*24*] (lognormal, log-series and power law, see supporting materials 6).

We now shift to examine the capability of NNTB to explain empirically observed patterns of population and community dynamics. To that end, we compare predictions of NNTB with those of UNTB using the dataset of the Barro Colorado Island (BCI), a long-term dataset of tropical forest dynamics that has been used extensively for tests of UNTB and its extensions [*3, 4, 13, 14*]. We focus on three fundamental patterns: the distribution of population fluctuations, the scaling of the magnitude of such fluctuations with initial population size, and the decay of compositional similarity over time (see supporting materials 3-5 for details).

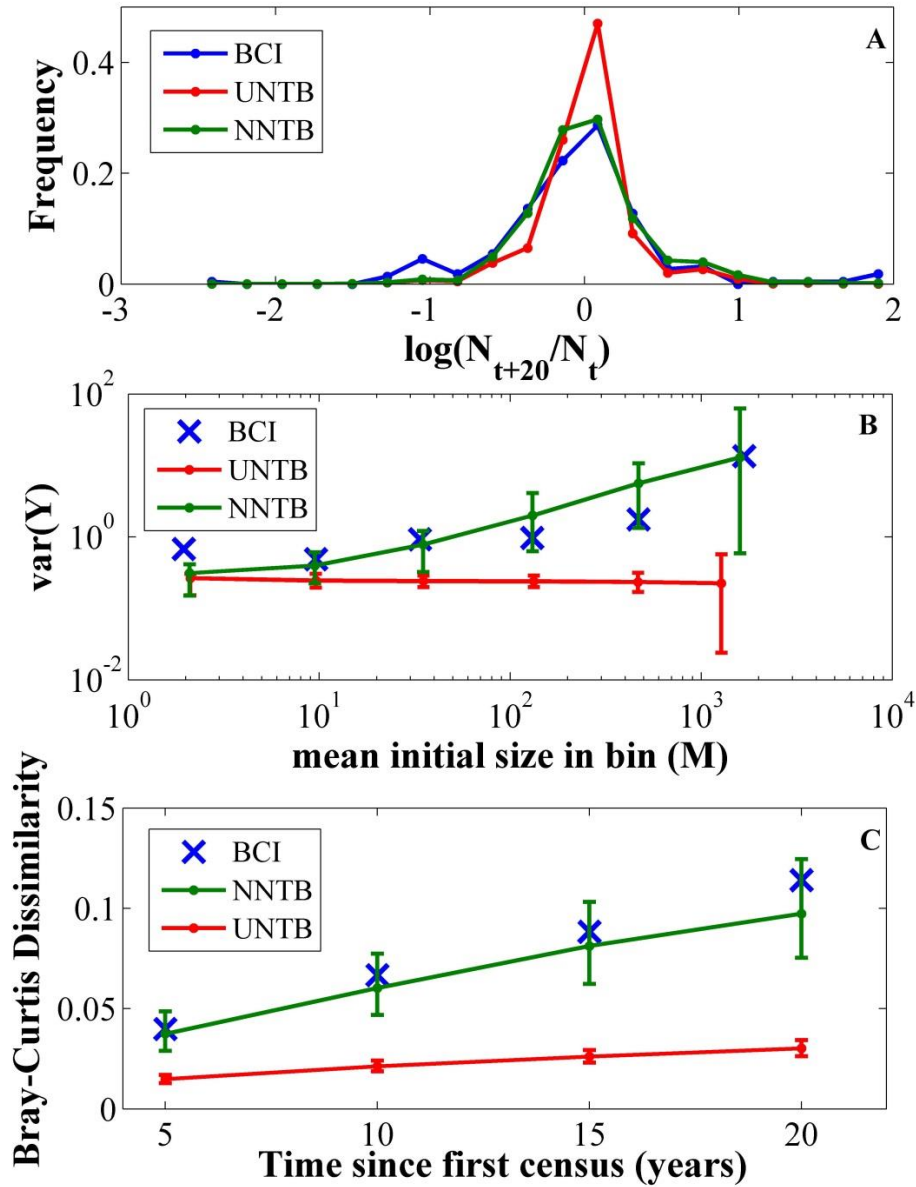

*Figure 2. A comparison of the capabilities of NNTB and UNTB to explain empirically observed patterns of population and community dynamics using data from the BCI forest. **A** presents the empirical distribution of 20 year fluctuations of tree populations in the BCI forest, along with the predictions of UNTB and GNTB (note the logarithmic scale). **B** compares the predictions of UNTB and NNTB regarding the scaling of normalized variance of 5 year fluctuations (var(Y)), with mean initial population size. The mean is calculated for logarithmic bins of base 4: 1-4, 5-16, .... **C** portrays the growth of community self-dissimilarity with time. Bray-Curtis dissimilarity was calculated between each BCI census that was conducted in the*

*period 1990-2005, and the initial census of 1985. Error bars in **B** and **C** represent 95% confidence intervals based on 1000 simulations (see supporting materials 4-5).*

Figure 2A presents a histogram of the empirically observed fluctuations in the BCI forest over 20 years [1985-2005] and the corresponding predictions of UNTB and NNTB. Clearly, NNTB provides a much better fit to the data. Moreover, the deviation of the empirical distribution from NNTB is not significant, while UNTB is ruled out (P=0.09 for NNTB, P<0.0001 for UNTB, see supporting materials 4-5). Analysis of smaller-scale fluctuations (5 years) reveals even more pronounced differences (P = 0.177 vs. P < 0.0001, respectively, Supplementary Figure 2).

To analyze the scaling of population fluctuations with population size, we normalized the observed fluctuations by the square root of initial population size as:

(2) $$Y_{\Delta t} = \frac{N_{\Delta t} - N_0}{\sqrt{N_0}}$$

The advantage of this approach is that under UNTB, the variance of the normalized fluctuations (*Y*) is independent of initial population size [*10*]. As can be seen in Fig. 2B, the empirical data show a strong and highly significant deviation from the UNTB prediction, but do not deviate from the 95% confidence intervals of the NNTB prediction except for the smallest populations. We attribute the latter deviation to the fact that both UNTB and NNTB assume a Moran birth-death process in which only a single individual dies each time step [*3*]. It is not unreasonable to assume that real populations may show a larger magnitude of demographic stochasticity.

Thus, NNTB provides better fit than UNTB to two different patterns of population dynamics. Importantly, both patterns were predicted using exactly the same parameter values, supporting the robustness of the results. Moving up to the community level, we compared the capabilities of NNTB and UNTB to predict the increase in community self-dissimilarity through time [*26*]. This comparison was performed using the same parameter values applied in the analyses of the population-level patterns (Fig. 2A, B). As before, NNTB offers a much better fit to the empirical data (Fig. 2C).

Finally we asked whether NNTB preserves the ability of UNTB to account for the distribution of species abundance in the BCI data. As can be seen in Figure 3, despite some differences, the explanatory power of NNTB and UNTB are rather similar. Qualitatively similar results are obtained if the two models are parameterized using the same values of *θ* and *m* obtained by Etienne [5] based on the exact sampling formula of UNTB (see figure S3). Thus, incorporating environmental stochasticity into Hubbell's neutral theory considerably improves its ability to explain patterns of population and community dynamics, while keeping its explanatory power with respect to patterns of commonness, rarity and richness. Although the inclusion of environmental stochasticity requires two additional parameters (magnitude of fitness differences and scale of temporal correlation in fitness of individual species), it is still a minimalistic model, much simpler than niche theories that require specification of the mutual effect of every pair of species [2].

We attribute the advantage of the newly presented NNTB over UNTB to the dual role played by environmental stochasticity: it is both a destabilizing force, thereby increasing the magnitude of temporal fluctuations in the abundance of common species, and a stabilizing force that promotes coexistence (through a storage effect). Interestingly, a recent analysis of the BCI data has found evidence for a significant storage effect [20]. It is also important to note that the storage effect predicted by NNTB reflects the fact that environmental stochasticity only affects recruitment rates, with no effect on mortality [18]. For example, a neutral model in which pairs of individuals are chosen at random and the identity of the 'winner' is determined based on the relative fitness of the two species does not produce storage effect [22].

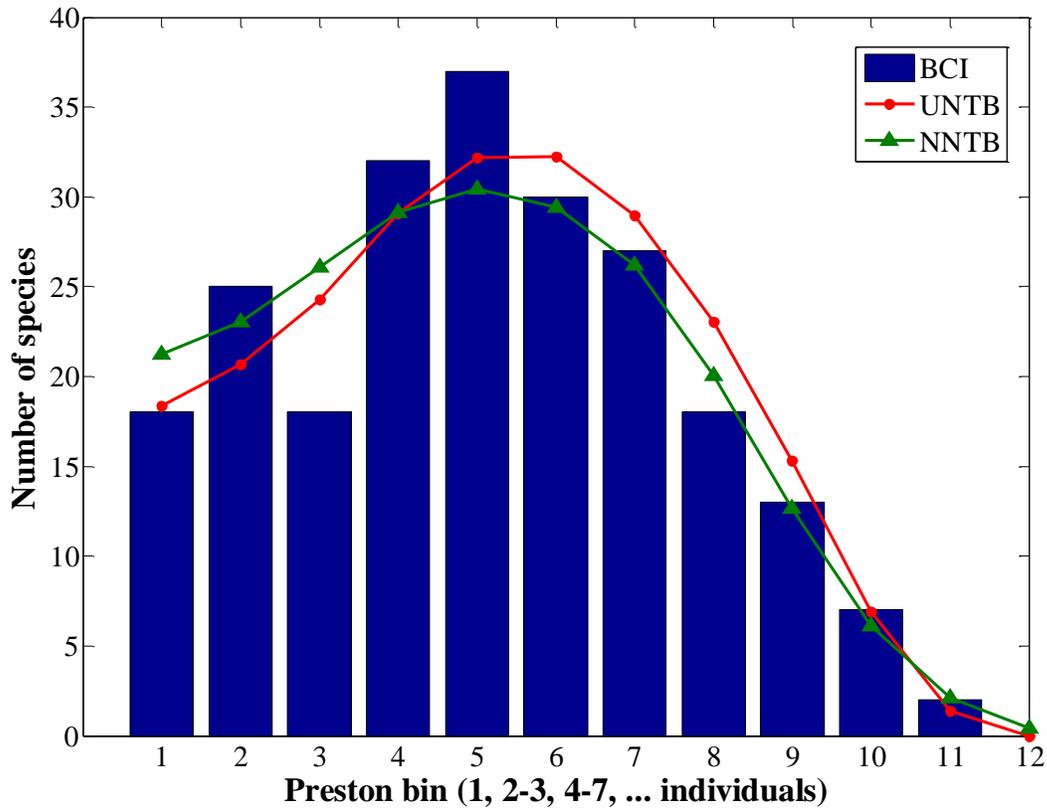

*Figure 3. Preston diagram of the species abundance distribution in the BCI forest and predictions based on UNTB and NNTB.* *The binning used is 1, 2-3, 4-7, ..., calculated for the 1995 census. $R^2$ for UNTB is 0.9301 and for GNTB 0.9265. The corresponding values of the Kolmogorov-Smirnov (K-S) statistic are 0.0571 and 0.0597, respectively.*

It has previously been noted that tests of ecological theories focusing on static patterns are rather weak because different processes may generate similar patterns [*7, 27*]. From this perspective, focusing on dynamic patterns is an advantage, particularly if several different patterns can be accounted for *simultaneously* by the same model *and* the same parameterization. Nevertheless, most tests of ecological theories still focus on static patterns (though see [*8, 10, 12, 28*]). Actually, we are aware of only one previous attempt to test a community dynamics model by fitting the same model with the same parameters to a set of static and dynamic patterns (Allen & Savage 2007) [*29*]. These authors demonstrated that the unrealistic extinction times predicted by UNTB may be resolved by increasing the magnitude of demographic stochasticity (note that their model does not explicitly incorporate environmental stochasticity) [*8, 16*].

Since the publication of UNTB, many attempts were made to resolve the "Niche-Neutrality debate" by incorporating deterministic niche forces with demographic stochasticity [*6, 7*]. Here we propose a conceptually different approach: instead of adding deterministic niche forces to UNTB we add stochastic niche forces while keeping the community neutral. Our results suggest that this minimalistic and general approach might be highly effective in explaining fundamental static and dynamic characteristics of ecological communities.

**Acknowledgements:** We thank C. H. Flather for useful comments on the manuscript. This work was supported by the Israeli Ministry of Science and Technology TASHTIOT program and by the Israeli Science Foundation grant no. 454/11 and BIKURA grant no. 1026/11. MK was supported by a grant from the Advanced School for Environmental Studies at HUJ. The BCI forest dynamics research project was made possible by National Science Foundation grants to S. P. Hubbell (DEB-0640386, DEB-0425651, DEB-0346488, DEB-0129874, DEB-00753102, DEB-9909347, DEB-9615226, DEB-9405933, DEB-9221033, DEB-9100058, DEB-8906869, DEB-8605042, DEB-8206992, and DEB-7922197); support from the Center for Tropical Forest Science, the Smithsonian Tropical Research Institute, the John D. and Catherine T. MacArthur Foundation, the Mellon Foundation, the Small World Institute Fund, and numerous private individuals; and the hard work of more than 100 people from 10 countries over the past 2 decades. The plot project is part of the Center for Tropical Forest Science, a global network of large-scale demographic tree plots.

# Supporting materials

## supporting figures

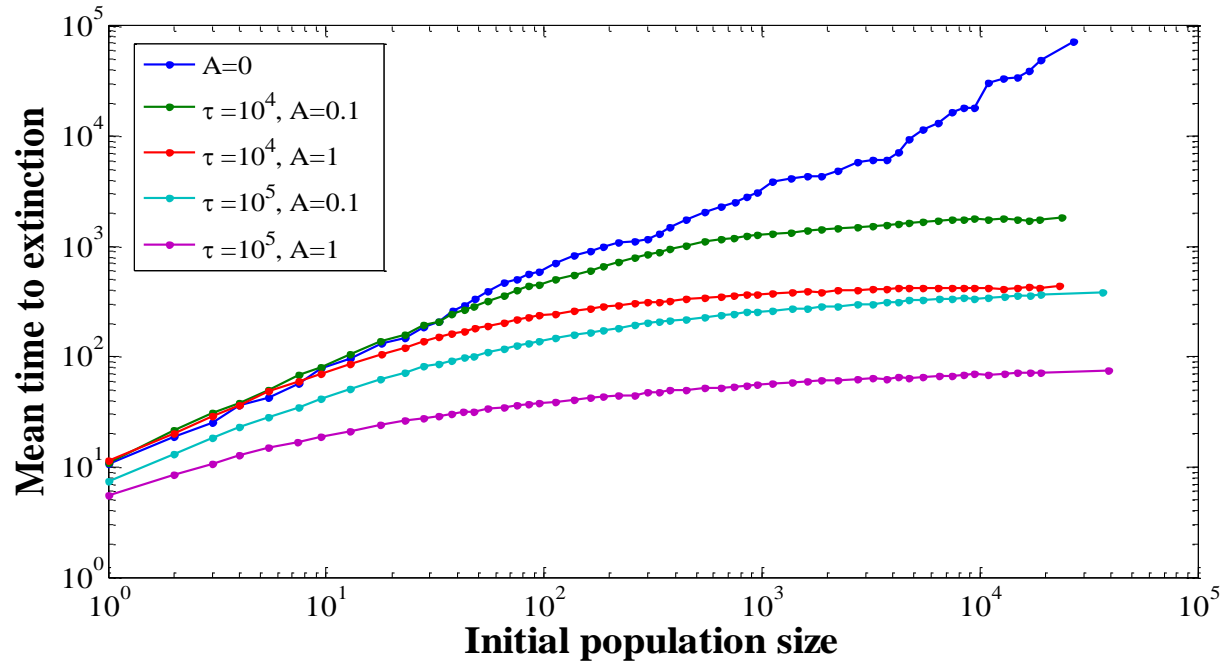

**Figure S1. Mean extinction times in simulations of a regional community with various parameters of environmental stochasticity**. For all five cases, $J_m=10^5$ and $\mu=10^{-4}$. Note that unlike figure 1 in the main text, under these parameters environmental stochasticity does not seem to enhance the longevity of small populations. This is probably because the higher speciation rates here (in contrast with $\mu=10^{-5}$ in figure 1 in the main text) lead to higher species richness, and such increase may weaken the strength of the storage effect[*19*]. See supporting material 1 for more details.

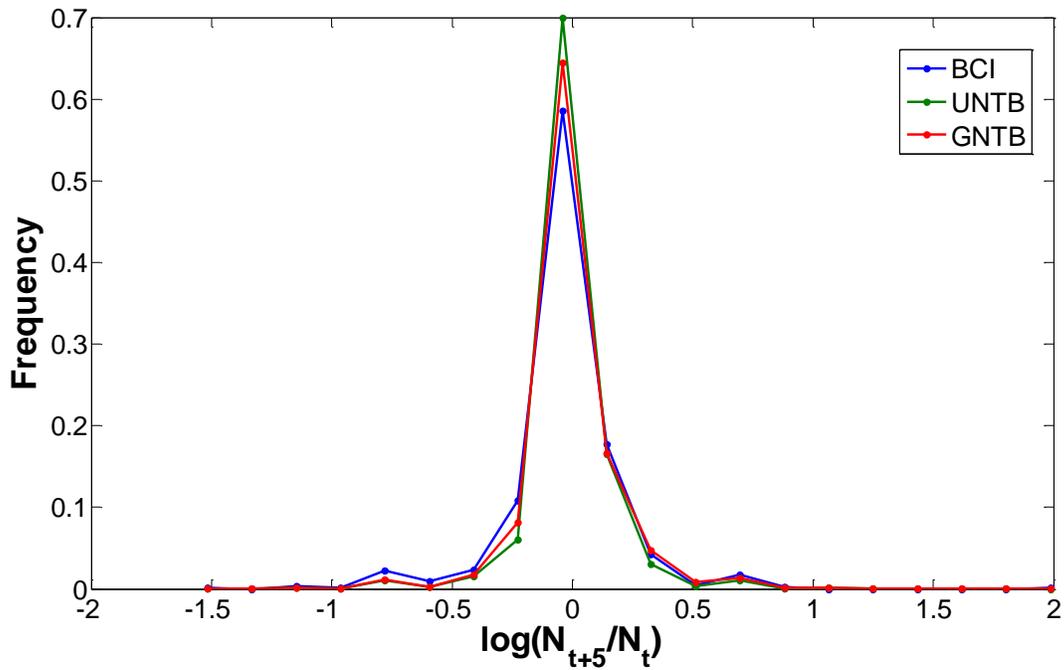

**Figure S2. Empirical distributions of 5 years fluctuations in population size of trees in the BCI forest and corresponding predictions based on NNTB and UNTB.** NNTB shows a much better fit to the data than UNTB when compared using both the Kolmogorov-Smirnov (K-S) (0.055 vs. 0.1070) and AIC (19240 vs. 19304) statistics. The deviation of the empirical distribution from the NNTB prediction is not significant (P=0.177) while the deviation from the UNTB prediction is highly significant (P<0.0001). See supporting material 5 for details.

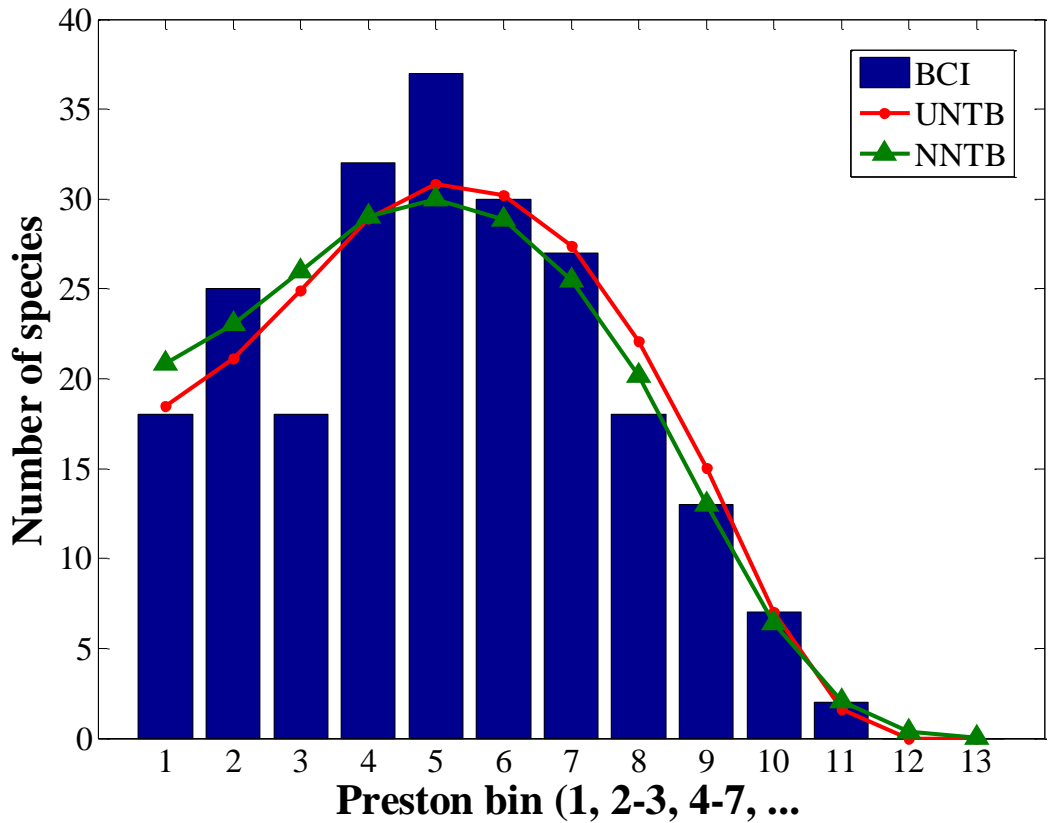

**Figure S3. Species abundance distribution in the BCI forest compared with theoretical predictions of NNTB and UNTB in which the values of *θ* and *m* were calculated using the exact sampling formula for UNTB (θ=47.8 and m=0.098).** For UNTB, R2 is 0.9298 and K-S is 0.0510. For NNTB, R2 is 0.9241 and K-S is 0.0601.

## Supporting material 1: Details of figure 1 in the main text

In Figure 1 we use the NNTB framework to analyze the effect of environmental stochasticity on coexistence, extinction time, and robustness to fitness differences. Here we bring the technical details of the simulations used in this figure. Panels A and B demonstrate the storage effect generated by introducing environmental stochasticity to the model. Local communities were simulated with J=1000, m=$10^{-3}$ and a regional species pool of 10 equally abundant species. Abundance through time plots of three typical species are depicted. In A we simulated a community with only demographic stochasticity (A=0), while in B there is also environmental stochasticity with A=1 and τ=100. The stabilization is apparent in the long run, as species attracted to an average size of *J/S* (~100 individuals).

Important details about panel C are brought in supporting material 2.

In panels D-F we examined the robustness of NNTB to differences in the mean fitness of species. The extreme sensitivity of the metacommunity to small deviations from perfect neutrality is one of the strongest theoretical arguments against UNTB. In UNTB, a slight fitness advantage of a single species over all other species is sufficient to completely alter the SAD, leading to the almost complete dominance of that species [*22, 23*]. It has previously been shown that incorporating density dependent stabilization into the model can mitigate this effect [*30*]. In our model, environmental stochasticity induces density dependent stabilization through a storage effect, leading to a similar result, although as an emergent property of the underlying dynamics.

To demonstrate that we run simulations of local scale NNTB, assuming immigration is very small (m=$10^{-3}$) and that the regional pool consists of 2000 equally abundant species with mean fitness of 10. This leads to dynamics resembling that of NNTB at a regional scale. We further introduced a single species with an advantage γ in mean fitness (γ = mean fitness of the added species minus mean fitness of all other species). For all cases in figure 1E-F, J=$2 \cdot 10^4$, m=$10^{-3}$ and τ=$2 \cdot 10^3$ turns. The simulations were run for $10^4$ generations of equilibration, and then sampled every 10 generations for 2000 generations.

The results show that under pure demographic stochasticity (A=0, figure 1D in main text), any level of competitive superiority leads to a collapse of the SAD into a single superior species (15$^{th}$ bin) and many rare species arriving via immigration or speciation. If environmental stochasticity is present (A=0.5 for figure 1E, A=3 for figure 1f in main text), the storage effect may buffer this asymmetry, but only up to some point. If competitive asymmetry is too strong ($\gamma$=1 under A=0.5, fig 1E), the SAD collapses again. Higher levels of environmental stochasticity can buffer higher degrees of competitive asymmetry, therefore recovering a neutral-like community (A=1 in fig 1F).

## Supporting material 2: Calculating mean time to extinction

In order to evaluate the ability of NNTB to resolve the apparent bias of UNTB in predicting mean time to extinction [*8, 31*] we examined the dependence of mean time to extinction on initial population size under several parameter regimes (see figure 1 in main text and figure S1).

Distributions of time to extinction tend to be very wide, requiring considerable computational effort (or analytical methods) to get reliable estimates of the mean. To overcome this difficulty, we have implemented the following procedure:

First, we run simulations for $10^4$ generations, where a generation is defined as J elementary steps (a step is a death of one tree and the reproduction of another). After this period we assume that the simulations equilibrated, and monitor the next $1.2 \cdot 10^5$ generations. For every population during the first $2 \cdot 10^4$ generations after equilibration, initial size was recorded, as well as the time it took the population to become extinct. The same population was often used and recorded multiple times, each time with different size and different time remaining till its extinction. The results were binned by initial population size in bins of increasing width and averaged.

In all the simulations $J_m$ was $10^5$. The parameter regime of pure demographic stochasticity with $\mu=10^{-5}$ led to very noisy results. For this regime we averaged the results over 5 independent runs and doubled all the times mentioned above. Still, some large populations did not become extinct by the end of the simulation (and were assigned the maximum time, over $10^5$ generations), making the last points in figure 1 an underestimate. For all other regimes, all the populations that were examined in the initial period (over 950 populations per bin, usually much more) had become extinct by the end of the simulation.

## Supporting material3: Calibrating time in the models

A major goal of this work is to examine the ability of neutral models to explain empirically observed dynamics of ecological communities. To achieve this it is important to calibrate time in the model correctly, that is, to find a transformation of time steps in the model into years in the empirically observed data. In this study we focused on the abundance of trees >10 DBH that were censused every five years in the period 1985-2005 in Barro Colorado Island, Panama [*32-34*]. Therefore, we had to determine the number of turns in the model (each turn consists of a mortality event followed by a recruitment event) that correspond to five years. We examined the individual level data, and found that in this period, 2459 trees on average were recruited from census to census, while 2439 perished, indicating 2449 observed events on average. However, a tree may be recruited and then die without being censused at all, and such "hidden events" must also be taken into consideration. We ran simulations of a birth-death process with J=21,500 individuals (as in BCI on average) and examined how many observed events correspond to a given number of true events. We found that 2449 observed events correspond to 2600 events, which was used as the sampling frequency for both UNTB and NNTB in the dynamic analyses (see figure 2 in main text for results).

Another important task is calibrating $\tau$ in NNTB, which is the frequency of redrawing fitness in the model. We did this by directly examining the empirically observed persistence of trends, taking into account the time series of populations above an initial size of 50. It turns out that, if a population had increased/decreased from a census to a second census, it had a 0.75 probability to keep increasing/decreasing from the second census to a third census, on average. Assuming lack of memory and symmetry, this indicates that in a period of 5 years a population has 0.5 probability of changing its relative fitness ("tossing the die" of fitness). Therefore, the expected time until the next time the population redraws its fitness is 10 years, which was used as $\tau$ for all analyses of NNTB.

## Supporting material 4: Fitting m, θ and A

To allow a transparent comparison between UNTB and NNTB, we assumed in both cases that the regional species pool has Fisher Log-Series distribution of abundances. We made several attempts to numerically calculate the likelihood of species abundance distributions under UNTB, but encountered systematic biases. Therefore, we used a somewhat crude but robust method for fitting the SAD, by maximizing the $R^2$ of the theoretical Preston Plot with respect to the 1995 census in BCI. We used a binning of 1, 2-3, 4-7, ... individuals per bin, which is the most recommended in the literature [35]. This fitting method inherently takes into consideration both the absolute number of species and the fact that the species are not independent, which are desired properties.

For fitting an SAD we ran a simulation with J=21,500 (as in the BCI on average) for 2000 generations (J turns each) for equilibration. We then sampled the simulation 4,001 times with an interval of 10 generations between consecutive samples, and calculated the average Preston Plot for all the samples.

We first fitted *m* and *θ* for UNTB. We found the best parameters to be m=0.08 and θ=51, close to the parameters obtained using the exact sampling formula [5]. Using these parameters, we proceeded to the fitting of the parameter A, the variance of the fitness distribution in NNTB. Note that we did not need to fit *τ* since its value was obtained from an independent analysis of the empirical data (see supporting material 3).

The parameter *A* controls the amplitude (variance) and the scaling of fluctuations. For this reason, we tried to find the optimal *A* simultaneously for three patterns: the distribution of population fluctuations in both 20 years interval (figure 2A in the main text), and 5 years interval (figure S2), and the scaling of 5 year fluctuation variance with initial size (figure 2B in the main text).

To quantify the fluctuations under different parameter regimes, simulations were run with J=21,500, the determined *θ*, *m*, *τ* and sampling frequency, as well as the required *A*, allowing 2000 generations for the system to equilibrate, and then using data from another 3000 generations. The goodness of fit to the distribution of empirical fluctuations was assessed using the Kolmogorov-Smirnov statistic (K-S), which we tried to minimize [36]. The goodness of fit to the fluctuation scaling was quantified as the $R^2$ of the var(*Y*) vs. *M*. diagram (see figure 2B)[10].

We found that for 20 year fluctuations, K-S was monotonously decreasing with A. For 5 year fluctuations, K-S was optimized at A=1. The fit of the scaling was optimized for A=0.6. We therefore use A=0.8 in our predictions of for all empirical patterns as presented in figure 2 and figure 3 in the main text.

To check consistency, we have tried to use other values of θ and m (rather than the values theta=51 and m=0.08 that yield the best fit for the UNTB SADs) for NNTB with A=0.8 and τ=10 years. Surprisingly, we found that the optimal values of theta and m were the same as for UNTB.

It is important to note that since A was fitted to the fluctuations and not for the SAD, it can be considered a constraint imposed on the fitting of NNTB to the SAD. Therefore, the good agreement of the SAD predicted by NNTB with the empirically observed SAD (see figure 3 in main text and figure S3) is highly non-trivial!

Finally, it should be noted that since we did not directly and simultaneously fitted our models to the different patterns, our results should be considered conservative, and there is probably a combination of parameters that better fits the data.

## Supporting material 5: Goodness of fit and significance in fluctuation analyses

Our main measure for the fit of UNTB and NNTB to empirically observed distributions of fluctuations and species abundance (figs. 2A and 3 in the main text, figs. S1 and S3) was the Kolmogorov-Smirnov statistic (K-S)[36]. We used this statistic also for testing whether the deviation between the observed and predicted distributions is statistically significant. For the latter analysis we compared the observed K-S of the empirical fluctuations with respect to the model to a theoretical distribution of K-S under the model, which was found in the following way: We ran 1000 simulations and sampled each only 5 times with 5 year intervals, and twice with a 20 years interval (as in the data). We calculated the K-S for the distribution of fluctuations in these short simulations to the with regards to long simulation of 3000 sampled generations (see supporting material 4). The comparison of the K-S of the empirical data to the theoretical distribution of K-S under the models gave us the significance for these models.

These 1000 short simulations were also used to calculate the error bars in figure 2B of the main text. The scaling of fluctuations was calculated (using the same binning as the empirical data) for each of these short simulations, and for each bin the error bars were estimated using the central 95% results. The 1000 short simulations were also used to calculate the mean growth of dissimilarity with time, and the 95% confidence intervals were constructed in the same manner (figure 3 in main text).

To evaluate whether the improvement in the goodness of fit of the fluctuations distribution may only be due to the extra fitted parameter *A*, we calculated the Akaike Information Criterion (AIC) of the fluctuations distribution for both NNTB and UNTB. The likelihood of a specific fluctuation was considered as the relative frequency of this fluctuation in the long simulation. <7% of the empirical fluctuations were not observed in the long simulation, and their likelihood was considered to be 1/*L*, where *L* is the length of the empirical distribution in the long simulation. The resulting AIC should be taken with a grain of salt, because it assumed independence of the fluctuations, and because of the treatment of non-observed fluctuations.

## Supporting material 6: Regional-scale SADs

UNTB predicts a log-series distribution of species abundance at the regional scale. However, empirical data show that other distributions can be found as well (see for example the SAD of Breeding Birds in the UK presented in Fig. 2.6 of Hubbell 2001[*3*]). A recent review of over 500 empirical datasets shows that natural communities exhibit three main types of SADs: Log-normal, Log-series, and Power-law (Geometric series)[*24*]. Clearly, accounting for such variable distributions is a crucial challenge for any general theory of biodiversity.

In figure S4 we show that NNTB can account for all three distributions under various parameter regimes in a predictable manner. The core feature that provides this versatility is the dual role played by environmental stochasticity. It destabilizes populations and causes fluctuations that lead to un-evenness of species and even extinctions, but at the same time it induces stability through the storage effect. If $\tau$, the scale of temporal correlation in fitness, is long (compared to the generation time), the storage effect becomes less effective [*19*] and yields a power-law SAD, as predicted by Kessler and Shnerb [*22*] for a theory with environmental noise but without storage effect. In the case of short $\tau$ and strong stochasticity (large *A*) the storage effect is strong and the SAD is peaked around $J_m/S$, where *S* is the number of species, resembling a left skewed Log-normal (figure S3C, F). Intermediate cases resemble a log-series (figure 3SB and figure 1 E-F in main text), as well as the case of A=0, when NNTB is reduced to UNTB.

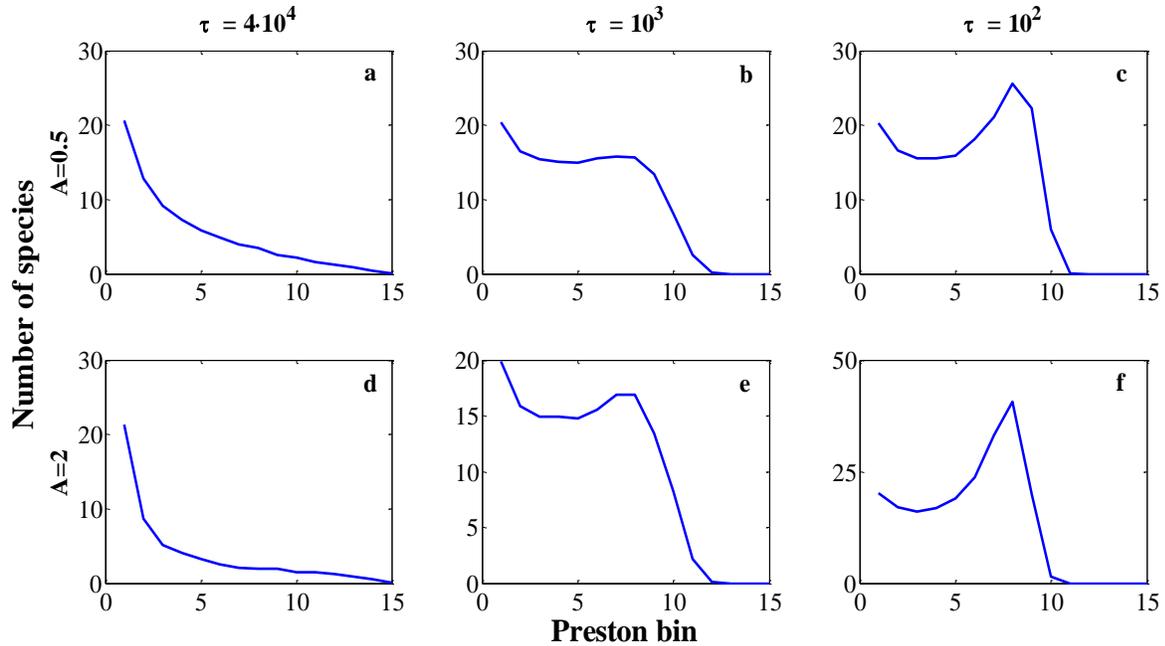

**Figure S4 | Regional SADs under NNTB.** For all the cases considered here, $J_m = 2 \cdot 10^4$ and $\mu = 10^{-3}$, while $\tau$ and A are altered. The simulations were run for $5 \cdot 10^3$ generations for equilibration, and then sampled every 10 generations for 2000 generations. The average number of species in every preston bin (1,2-3,4-7… individuals per bin) is presented.

It is important to note that the form of the SAD is also sensitive to $J_m$ and $\mu$, since an increase in the number of species weakens the storage effect [*19*].